\documentclass[twocolumn,prl,showpacs,superscriptaddress,preprintnumbers,amsmath,amssymb,floatfix]{revtex4}
\usepackage[dvips]{graphicx}
\usepackage{dcolumn}
\usepackage{bm}
\usepackage{color}
\begin{document}
\newcommand{\Ru}{La$_4$Ru$_2$O$_{10}$}

\title{Orbitally driven spin-singlet dimerization in $S$=1 La$_4$Ru$_2$O$_{10}$}

\author{Hua Wu}
  \affiliation{II. Physikalisches Institut, Universit\"{a}t zu K\"{o}ln, Z\"{u}lpicher Str. 77, 50937 K\"{o}ln, Germany}
\author{Z.~Hu}
  \affiliation{II. Physikalisches Institut, Universit\"{a}t zu K\"{o}ln, Z\"{u}lpicher Str. 77, 50937 K\"{o}ln, Germany}
\author{T.~Burnus}
  \affiliation{II. Physikalisches Institut, Universit\"{a}t zu K\"{o}ln, Z\"{u}lpicher Str. 77, 50937 K\"{o}ln, Germany}
\author{J.~D.~Denlinger}
  \affiliation{Advanced Light Source, Lawrence Berkeley National Laboratory, Berkeley, CA 94720, USA}
\author{P.~G.~Khalifah}
  \affiliation{Department of Chemistry, University of Massachusetts, Amherst, MA 01003, USA}
\author{D.~Mandrus}
  \affiliation{Condensed Matter Sciences Division, Oak Ridge National Laboratory, Oak Ridge, TN 37831, USA}
\author{L.-Y. Jang}
  \affiliation{National Synchrotron Radiation Research Center, 101 Hsin-Ann Road, Hsinchu 30077, Taiwan}
\author{H.~H.~Hsieh}
  \affiliation{Chung Cheng Institute of Technology, National Defense University, Taoyuan 335, Taiwan}
\author{A.~Tanaka}
  \affiliation{Department of Quantum Matter, ADSM, Hiroshima University, Higashi-Hiroshima 739-8530, Japan}
\author{K.~S.~Liang}
  \affiliation{National Synchrotron Radiation Research Center, 101 Hsin-Ann Road, Hsinchu 30077, Taiwan}
\author{J.~W.~Allen}
  \affiliation{Randall Laboratory of Physics, University of Michigan, Ann Arbor, MI 48109, USA}
\author{R.~J.~Cava}
  \affiliation{Department of Chemistry, Princeton University, Princeton, NJ 08540, USA}
\author{D.~I.~Khomskii}
  \affiliation{II. Physikalisches Institut, Universit\"{a}t zu K\"{o}ln, Z\"{u}lpicher Str. 77, 50937 K\"{o}ln, Germany}
\author{L.~H.~Tjeng}
  \affiliation{II. Physikalisches Institut, Universit\"{a}t zu K\"{o}ln, Z\"{u}lpicher Str. 77, 50937 K\"{o}ln, Germany}

\date{\today}

\begin{abstract}
Using x-ray absorption spectroscopy at the Ru-$L_{2,3}$ edge we
reveal that the Ru$^{4+}$ ions remain in the $S$=1 spin state
across the rare $4d$-orbital ordering transition and spin-gap
formation. We find using local spin density approximation +
Hubbard U (LSDA+U) band structure calculations that the crystal
fields in the low temperature phase are not strong enough to
stabilize the $S$=0 state. Instead, we identify a distinct
orbital ordering with a significant anisotropy of the
antiferromagnetic exchange couplings. We conclude that {\Ru}
appears to be a novel material in which the orbital physics drives
the formation of spin-singlet dimers in a quasi 2-dimensional
$S$=1 system.
\end{abstract}


\maketitle

One of the most intriguing aspects of transition metal materials
is the wide variety and richness of their physical properties
\cite{Imada98}. Although conceptually clean and beautiful,
theoretical simplifications in terms of a Heisenberg model or a
single band Hubbard model turn out to be inadequate
\cite{Tokura00}. It now becomes more and more clear that a full
identification of the relevant orbital and spin degrees of
freedom of the ions involved is needed to understand, for
instance, the colossal magneto-resistance behavior in the
manganates \cite{Ramirez97,Khomskii97,Mizokawa95}, magnetization
reversals and metal-insulator transitions in early transition
metal oxides \cite{Ren98,Blake01,Ulrich03,Park00,Haverkort05}, as
well as the formation of spin-gaps in non-1-dimensional
$S$=$\frac{1}{2}$ systems
\cite{Isobe02,Radaelli02,Schmidt04,Khomskii05}.

Very recently Khalifah \textit{et al.} \cite{Khalifah02}
synthesized the semiconducting quasi-2-dimensional {\Ru} compound
and discovered that this system undergoes a strong first-order
structural transition at $T_s$=160 K (see Fig. 1), accompanied by
a rare $4d$-orbital ordering and spin-gap opening. Their
interpretation of these phenomena was that the Ru$^{4+}$ ion
transforms from the usual $t_{2g\uparrow}^3t_{2g\downarrow}^1$
low-spin state with $S$=1 to an
$t_{2g\uparrow}^2t_{2g\downarrow}^2$ `ultra-low' spin state with
$S$=0, caused by a sufficiently strong crystal-field splitting
(CFS) of the Ru $4d$-$t_{2g}$ levels due to the lattice
distortion below $T_s$.

\begin{figure}[ht!]
    \centering
    \includegraphics[width=7cm]{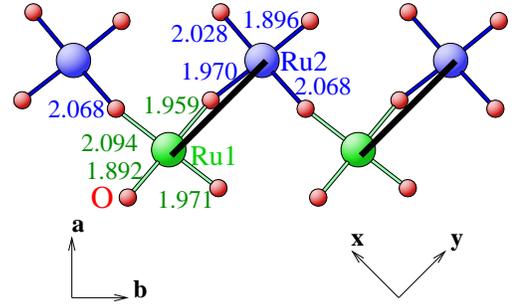}
    \caption{(Colour online) The main block of the low-temperature (20 K) triclinic
     crystal structure of {\Ru} shown in the crystallographic $ab$ plane:
     The 2-dimensional Ru-O network consisting of distorted RuO$_{6}$
     octahedra extends along the $c$-axis (not shown, pointing into paper plane)
     and $b$-axis with corrugation.
     The Ru-O bond-lengths are marked in unit of \AA, and those
     out of the $ab$-plane (not shown) are 2.046 and 2.057 \AA\ for Ru1, and
     2.046 and 2.082 \AA\ for Ru2. {\it The spin-singlet dimers are marked by
     black solid bars.} The local orthogonal coordinate
     system ($xyz$) is used in our band calculations with $z$
     parallel to $c$ and $y$ ($x$) along the short (long) Ru1-Ru2 direction. The
     high-temperature monoclinic structure (not shown) above $T_s$=160
     K has equal Ru-Ru distances along the $x$ and $y$ directions: Ru1
     and Ru2 are equivalent.}
     \label{fig1}
\end{figure}

However, already soon after that, it was also hypothesized by
Khalifah \textit{et al.} based on unpublished standard band
structure calculations that a chemical bond may
be formed associated with the orbital ordering. Here we report on
an x-ray absorption spectroscopy (XAS) study in which we reveal
that the Ru$^{4+}$ ions remain in the $S$=1 spin state across
$T_s$. This directly points to the possibility that {\Ru} is in
fact a novel system in which the spin-gap opening is due to a
singlet dimer formation in a non-1-dimensional and
$S$$>$$\frac{1}{2}$ material. We find using LSDA+U band structure
calculations that the distinct orbital ordering involves a
significant anisotropy of the antiferromagnetic exchange
couplings, indicating indeed the formation of Ru($S$=1)-Ru($S$=1)
singlet dimers.

Floating zone crystals were grown in an NEC SC-M15HD image
furnace using rods with a 1:1 or 4:5 ratio of La:Ru which had
been pre-reacted and sintered in air at 1250$^{\circ}$C.  A small
(1-2 atm) overpressure of oxygen aided the growth, and the power
was dynamically increased during the run to compensate for
absorption by the copious amounts of evaporated Ru. Sizeable
crystals could only be obtained using seed crystals. X-ray
diffraction confirmed both the macroscopic phase purity and the
universal presence of two twin domains. The XAS measurements were
performed at the Taiwan NSRRC 15B beamline, equipped with a
double Si(111) crystal monochromator delivering photons from 2
keV and up. The spectra were recorded using the total electron
yield method in a chamber with a base pressure in the low
10$^{-10}$ mbar range. Clean sample areas were obtained by
cleaving the crystals \textit{in-situ}. The photon energy
resolution at the Ru $L_{2,3}$ edges ($h\nu$$\approx$2.9 keV) was
set at 0.6 eV. Strong polarization dependent O-$K$ XAS spectra
\cite{Denlinger05} verify the high sample quality.

\begin{figure}[ht!]
    \centering
    \includegraphics[width=8.5cm]{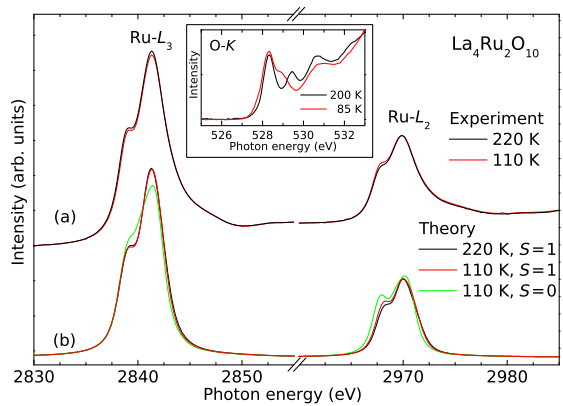}
    \vspace{-5mm}
    \caption{(Colour online). (a) Experimental Ru-$L_{2,3}$ XAS spectra
    of {\Ru} measured at 220 and 110 K,
    i.e. above and below $T_s$=160 K.
    (b) Theoretical simulations for the Ru$^{4+}$ ion in the $S$=1 high-temperature
    phase at 220 K, $S$=1 low-temperature phase at 110 K, and $S$=0.
    The inset shows O-$K$ spectra measured at 200 and 85 K from \cite{Denlinger05}.}
    \label{fig2}
\end{figure}

The top curves of Fig. 2 depict the Ru-$L_{2,3}$ XAS spectra of
{\Ru} taken at 220 K (black line) and 110 K (red line). The
spectral line shapes depend strongly on the multiplet structure
given by the Ru $4d$-$4d$ and $2p$-$4d$ Coulomb and exchange
interactions, as well as by the local CFS and the hybridization
with the O $2p$ ligands. Unique to XAS is that the dipole
selection rules are very effective in determining which of the
$2p^{5}4d^{n+1}$ final states can be reached and with what
intensity, starting from a particular $2p^{6}4d^{n}$ initial
state ($n$=4 for Ru$^{4+}$) \cite{deGroot94,Thole97}. This makes
the technique extremely sensitive to the quantum numbers of the
initial state \cite{Hu00,Hu04}.

The essence of the Ru-$L_{2,3}$ XAS spectra in Fig. 2 is that
there is only a very small change across $T_s$, suggesting that
the local electronic and spin state of the Ru$^{4+}$ ion in the
high temperature (HT) and low temperature (LT) phases are quite
similar. This seems surprising in view of the fact that we
observed considerable modifications in the O-$K$ XAS spectra in
going from 200 K (black line) to 85 K (red line) as shown in the
inset of Fig. 2 \cite{Denlinger05}. Since these O-$K$ spectra are
known to be sensitive to band structure effects \cite{deGroot94},
their modifications are fully consistent with the strong changes
in the crystal structure as seen in neutron diffraction
\cite{Khalifah02}, confirming once again the good quality of our
samples.

To extract quantitative information on the CFS and spin state from
the Ru spectra, we have performed simulations using the successful
configuration interaction cluster model
\cite{deGroot94,Thole97,Tanaka94}. The calculations have been
carried out for a RuO$_6$ cluster in the proper HT and LT local
coordination using the XTLS 8.3 program~\cite{Tanaka94}.
Parameters for the multipole part of the Coulomb interactions
were set standardly at 80\% of the Hartree-Fock values
\cite{Tanaka94}, while the monopole parts ($U_{dd}$, $U_{pd}$)
were taken from Ca$_2$RuO$_4$ \cite{Mizokawa0104,footnote1}. The
O $2p$ - Ru $4d$ charge transfer energy was estimated from LDA
calculations (see below), and the O $2p$ - Ru $4d$ transfer
integrals and their distance dependence from Harrison's relations
\cite{Harrison89}. The local CFS parameters are to be determined
from the comparison between the simulations and the experimental
spectra.

The bottom curves of Fig. 2 show the simulations for both the HT
and LT phases. We found optimal fits (black and red lines) if the
Ru-$4d$ $xz$ orbital is set at about 100-150 meV (HT) and 200-300
meV (LT) lower in energy than the essentially degenerate (within
50 meV) $yz$ and $xy$ orbitals \cite{parameters}. These numbers
refer to total energies calculated including the CFS and
covalency but without spin-orbit interaction. Important is that
the cluster calculation indeed finds the $S$=1 state for both the
HT and LT phases, which is a direct consequence of the fact that
in both phases the $xz$ orbital is essentially doubly occupied
while the $yz$ and $xy$ are each singly occupied. We also have
carried out calculations for the Ru ion in the artificial $S$=0
state by changing the CFS parameter such that the $xz$ orbital
lies above the degenerate $yz$ and $xy$ orbitals
\cite{parameters}. As shown in Fig. 2, the simulated spectrum
(green curve) disagrees with the experiment. So we can rule out
that the spin-gap opening is due to a local spin state transition
\cite{Khalifah02}.

To confirm our XAS-derived conclusions and, more importantly, to
get in-depth understanding of the nature of the spin-gap state
below $T_s$, we performed systematic LDA and LSDA+U band structure
calculations \cite{Anisimov91}, by using the full-potential
augmented plane waves plus local orbital method \cite{WIEN2k}. We
took the neutron crystal structure data at 20 K and 298 K
\cite{Khalifah02}. The muffin-tin sphere radii are
chosen to be 2.8, 2.0 and 1.5 Bohr for La, Ru and O atoms,
respectively. The cut-off energy of 13 Ryd is used for plane wave
expansion of interstitial wave functions, and 120 {\bf k}-mesh
for integrations over the Brillouin zone. U=3 eV and Hund's rule
exchange J$_H$=0.5 eV (U$_{eff}$=2.5 eV) are used for Ru $4d$
electrons, which are common for
ruthenates~\cite{Mizokawa0104,Fang04,footnote2}.

\begin{figure}[ht!]
\centering\includegraphics[width=8cm]{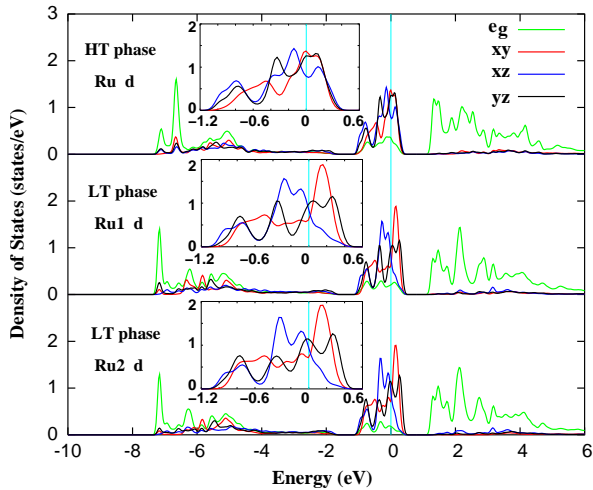}
 \caption{(Colour online). LDA density of states of the Ru $4d$
orbitals in the high-temperature (HT, upper panel) and
low-temperature (LT, middle and lower panels) phases.
The insets show a close up of the $t_{2g}$ ($xy$, $xz$, and $yz$)
states in the vicinity of the Fermi level set at 0 eV.}
 \label{fig3}
\end{figure}

Fig. 3 shows the Ru $4d$ density of states (DOS) calculated using
the LDA for the non-magnetic (NM) state. The inset shows a
close-up of the $t_{2g}$ levels and a calculation of the first
moments supports the XAS analysis: for both the HT and LT phases,
the $xz$ orbital lies lowest and the splitting $\Delta^{'}_{CF}$
between the higher lying $yz$ and $xy$ orbitals is less than 50
meV. It is the magnitude of this $\Delta^{'}_{CF}$ relative to
J$_H$ which determines the spin state of the Ru$^{4+}$ ion. To
make a crude estimate: the $S$=1 state
($xz^{\uparrow\downarrow}$$yz^{\uparrow}$$xy^{\uparrow}$) carries
the Hund's stabilization energy of 3$J_H$, whereas the $S$=0
state ($xz^{\uparrow\downarrow}$$yz^{\uparrow\downarrow}$) has a
total stabilization energy of 2$J_H$ plus $\Delta^{'}_{CF}$.
Assuming that 0.5 eV is a reasonable estimate for J$_H$, one must
expect that a $\Delta^{'}_{CF}$ of 0.05 eV is far from sufficient
to obtain the $S$=0 state. Obviously this is what the XAS
experiments have revealed. Moreover, our LDA calculations also
find that the ferromagnetic (FM) as well as the antiferromagnetic
(AF) solution are more stable than the NM one, giving further
support that the $S$=0 is unfavorable. This was also recently
found by GGA calculations of Eyert $et$ $al$~\cite{Eyert}.

\begin{figure*}[ht!]
\centering\includegraphics[width=16cm]{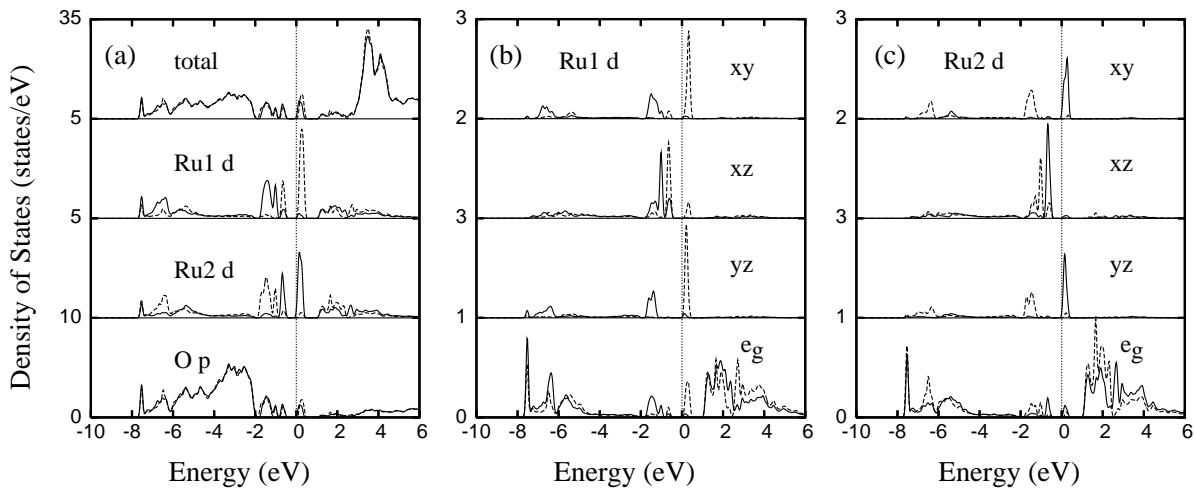}
 \caption{Density of states (DOS) of the {\Ru} in the antiferromagnetic
 low temperature phase calculated using LSDA+U.
 (a) Total DOS per formula unit, the $4d$ states of the two
 inequivalent Ru$^{4+}$ ions, and the $2p$ states of all the ten
 oxygens; (b) orbitally-resolved $4d$ states of Ru1 and (c) of Ru2.
 The solid (dashed) curves denote the up (down)-spin channels.}
 \label{fig4}
\end{figure*}

The LDA results as shown in Fig. 3 predict a metallic state for
both the HT and LT phases, and this is in strong disagreement
with the observed semiconducting behavior with a gap
of about 0.3 eV \cite{Khalifah02}. One should note however, that
the Fermi level is located in a narrow Ru $t_{2g}$ band with no
more than 1.5 eV width. This signals that modest electron
correlation effects at the Ru sites will already be able to turn
this material into a Mott insulator. We therefore set out to do
LSDA+U calculations for the LT phase and found that a band gap of
about 0.5 eV is indeed opened in the Ru $t_{2g}$ band as can be
seen from Fig. 4a. Within the LSDA+U mean-field approach, the
lowest state of this insulator is AF and is labeled as AF$_{xyz}$
in Table I to indicate the AF alignment with the nearest
neighbors along the $x$, $y$, and $z$ directions (Fig. 1). The
Ru$^{4+}$ spin moment inside the muffin-tin sphere is about 1.2
$\mu_B$, confirming the XAS result that the spin state is $S$=1
but not $S$=0.

It is important to look now at the orbital character of the
AF$_{xyz}$ solution. Fig. 4b and 4c show the orbitally resolved
DOS of the AF aligned Ru1 and Ru2 ions. One can clearly see for
each of the ions that the $xy$ and $yz$ orbitals (with the spins
parallel) are singly occupied while the $xz$ are doubly occupied.
This double occupation is due to the Ru-O bonds being elongated
along the $x$ direction, see Fig. 1. We thus find an orbital
ordered state which is different from the originally proposed
doubly occupied $xy$ and $xz$ ($S$=0) state \cite{Khalifah02}.

As a result, the half-filled $xy$ and $yz$ orbitals are
magnetically active. To explain the formation of the spin gap in
the LT phase, it is crucial to identify the relevant exchange
interactions in this system. We therefore have calculated other
magnetic configurations: the FM and two more types of AF
solutions, namely AF$_{xy}$ and AF$_{yz}$. The 2$\times${\Ru}
unit cell is used for all solutions, except for AF$_{yz}$ where
the 4$\times${\Ru} unit cell is taken with the doubling along the
$b$ direction (see Fig. 1). We also studied the NM solution, and
confirmed that this is much higher in energy than the AF$_{xyz}$,
by 775 meV per formula unit (Table I). The NM solution is
metallic, the FM half-metallic and all AF insulating. The
Ru$^{4+}$ spin moment is 1.2 $\pm$ 0.1 $\mu_B$ for all magnetic
solutions.

The relative energies of the different magnetic states allow us to
estimate the exchange constants $J_x$, $J_y$, and $J_z$ along the
$x$, $y$, and $z$ directions (Fig. 1), respectively. With the FM
being 130-150 meV higher in energy than the AF solutions, we thus
have very large AF exchange interactions in this system. As
listed in Table I, we find $J_x$=1.5 meV, $J_y$=65.5 meV and
$J_z$=4.5 meV. The significant anisotropy is related to the fact
that \textit{both} the $xy$ and $yz$ orbitals contribute to the
exchange coupling along the $y$ direction having the
\textit{short} Ru-O-Ru distance, while \textit{only} the $xy$
\textit{or} $yz$ orbital contributes to the $x$ \textit{or} $z$
direction, respectively, having the \textit{long} Ru-O-Ru
distance. Hence we can consider the LT phase of {\Ru} as
practically consisting of strongly coupled Ru-Ru dimers with weak
inter-dimer coupling, or at most as two-leg ladders with
$J_{rung}$=$J_y$=65.5 meV and $J_{leg}$=$J_z$=4.5 meV, with weak
interladder coupling of 1.5 meV. Those rather well isolated Ru-Ru
dimers or rungs will have the singlet ground state
\cite{footnote3}. This explains naturally the appearance of a
spin gap in the LT phase, which, according to our calculations
should be about 60 meV, in reasonable agreement with the
experimentally observed value of 40 meV~\cite{Khalifah02}. The
dimer character of the spin gap seems also to agree with the
results of the single-crystal neutron scattering \cite{Osborn}.

\begin{table}[h!]
 \caption{LSDA+U results for the total energy (per
formula unit and relative to the lowest solution) and band gap in
the low-temperature phase of {\Ru}, calculated for the
non-magnetic (NM), ferromagnetic (FM) and three types of
antiferromagnetic (AF) solutions. The exchange constants are
found to be $J_x$=1.5 meV, $J_y$=65.5 meV, and $J_z$=4.5 meV.}
\label{TableI}
\begin{tabular}
 {l|c|c|c} \hline
state&exchange& energy (meV) & gap (eV) \\ \hline
NM& - &775&-- \\
FM& $J_x$+$J_y$+2$J_z$&152&-- \\
AF$_{xyz}$& --$J_x$--$J_y$--2$J_z$&0&0.5 \\
AF$_{xy}$& --$J_x$--$J_y$+2$J_z$&18&0.3 \\
AF$_{yz}$& $J_x$--$J_y$--2$J_z$&3&0.4 \\
\hline
\end{tabular}
\end{table}

To summarize, XAS measurements revealed that the Ru$^{4+}$ ions
in {\Ru} remain in the $S$=1 spin state across the structural
phase transition and spin-gap formation. LSDA+U calculations
provided support for this finding and identified the distinct
orbital ordering accompanying the structural transition. Crucial
is that with LSDA+U we were able to estimate the inter-site
antiferromagnetic exchange interactions and found them to be
highly anisotropic. This brought us to the conclusion that the
spin-gap opening is due to the formation of Ru-Ru singlet dimers.
Such a transition is rather unusual since {\Ru} is a 2-dimensional
$S$=1 system; it is largely driven by orbital ordering which
amplifies the importance of orbital physics in correlated systems.

We acknowledge the NSRRC for the extremely stable beam. We thank
M. W. Haverkort for valuable discussions and Lucie Hamdan for her
skillful technical assistance. The research in Cologne is
supported by the Deutsche Forschungsgemeinschaft through SFB 608
and by the European project COMEPHS.



\begin{thebibliography}{99}


\bibitem{Imada98} M. Imada, A. Fujimori, and Y. Tokura,
 Rev. Mod. Phys. {\bf 70}, 1039 (1998).

\bibitem{Tokura00} Y. Tokura and N. Nagaosa,
 Science \textbf{288}, 462 (2000).

\bibitem{Ramirez97} A. P. Ramirez,
     J. Phys.: Conden. Matter {\bf 9}, 8171 (1997).

\bibitem{Khomskii97} D. I. Khomskii and G. A. Sawatzky,
 Solid State Commun. {\bf 102}, 87 (1997).

\bibitem{Mizokawa95} T. Mizokawa and A. Fujimori,
 Phys. Rev. B {\bf 51}, 12880 (1995);
 \textit{ibid.} {\bf 54}, 5368 (1996);
 \textit{ibid.} {\bf 56}, R493 (1997).

\bibitem{Ren98} Y. Ren \textit{et al.},
 Nature \textbf{396}, 441 (1998).

\bibitem{Blake01} G. R. Blake \textit{et al.},
 Phys. Rev. Lett. \textbf{87}, 245501 (2001).

\bibitem{Ulrich03} C. Ulrich \textit{et al.},
 Phys. Rev. Lett. \textbf{91}, 257202 (2003).

\bibitem{Park00} J.-H. Park \textit{et al.},
 Phys. Rev. B {\bf 61}, 11506 (2000).

\bibitem{Haverkort05} M.~W. Haverkort \textit{et al.},
 Phys. Rev. Lett. \textbf{95}, 196404 (2005).

\bibitem{Isobe02} M. Isobe \textit{et al.},
 J. Phys. Soc. Jpn. \textbf{71}, 1423 (2002).

\bibitem{Radaelli02} P.~G. Radaelli \textit{et al.},
 Nature \textbf{416}, 155 (2002).

\bibitem{Schmidt04} M. Schmidt \textit{et al.},
 Phys. Rev. Lett. \textbf{92}, 056402 (2004).

\bibitem{Khomskii05} D. I. Khomskii and T. Mizokawa,
 Phys. Rev. Lett. \textbf{94}, 156402 (2005).

\bibitem{Khalifah02} P. Khalifah \textit{et al.},
 Science \textbf{297}, 2237 (2002).

\bibitem{Denlinger05} J. D. Denlinger \textit{et al.},
 (unpublished).

\bibitem{deGroot94} F. M. F. de Groot,
 J. Electron Spectrosc. Relat. Phenom. {\bf 67}, 529 (1994).

\bibitem{Thole97} see the Theo Thole Memorial Issue,
 J. Electron Spectrosc. Relat. Phenom. {\bf 86}, 1 (1997).

\bibitem{Hu00} Z. Hu \textit{et al.},
 Phys. Rev. B \textbf{61}, 5262 (2000).

\bibitem{Hu04} Z. Hu \textit{et al.},
 Phys. Rev. Lett. \textbf{92}, 207402 (2004).

\bibitem{Tanaka94} A. Tanaka and T. Jo,
  J. Phys. Soc. Jpn. \textbf{63}, 2788 (1994).

\bibitem{Mizokawa0104} T. Mizokawa \textit{et al.},
 Phys. Rev. Lett. \textbf{87}, 077202 (2001);
 Phys. Rev. B \textbf{69}, 132410 (2004).

\bibitem{footnote1} Both La$_4$Ru$_2$O$_{10}$ and
Ca$_2$RuO$_4$ are 2-dimensional and have almost the same
octahedral $<$Ru$^{4+}$-O$>$ bond-lengths and similar $t_{2g}$
bandwidths of about 1.5 eV.

\bibitem{Harrison89} W. A. Harrison, \textit{Electronic Structure
 and the Properties of Solids} (Dover, New York, 1989).

\bibitem{parameters} Parameters for RuO$_{6}$ cluster [eV]:
$U_{dd}$=3.0, $U_{cd}$=2.0, $\Delta$=2.0, $pd\sigma$=--2.1 for
2.01\AA, 10$Dq$=1.9, $Dt$=0.04, $Du$=0.04, $Dv$=0.00, $\zeta$=
60\% of Hartree-Fock value; $Ds$=--0.03 ($S$=1,HT), --0.06
($S$=1,LT), +0.90 ($S$=0,LT).

\bibitem{Anisimov91} V. I. Anisimov \textit{et al.},
 Phys. Rev. B \textbf{48}, 16929 (1993).

\bibitem{WIEN2k} P. Blaha \textit{et al.},
 http://www.wien2k.at.

\bibitem{Fang04} Z. Fang, N. Nagaosa, and K. Terakura,
 Phys. Rev. B \textbf{69}, 045116 (2004).

\bibitem{footnote2} Using a larger Hubbard U=5 eV
does not change the spin-singlet dimer picture at all, but gives
a too large band gap of 1.1 eV.

\bibitem{Eyert} V. Eyert, S. G. Ebbinghaus, and T. Kopp,
cond-mat/0512409.

\bibitem{footnote3} Inclusion of the spin-orbit coupling does not 
affect our conclusions: the $xz$ orbital in the LT phase being 
lower than the $yz$ and $xy$ by the CFS of $\approx$300 meV together 
with the band formation make the weaker spin-orbit coupling
of $\approx$150 meV to be less operative. The exchange constants
change by less than 5 meV and the orbital moment is small, 
not more than about 0.2 $\mu_B$.

\bibitem{Osborn} R. Osborn, private communication.

\end{thebibliography}
\end{document}